\documentclass[twocolumn]{aastex631}
\shorttitle{Learn Spectroscopy From Colors}
\shortauthors{Jafariyazani et al.}
\graphicspath{{./}{figures/}}

\begin{document}

\title{Predicting the Spectroscopic Features of Galaxies by Applying Manifold Learning on Their Broad-Band Colors: Proof of Concept and Potential Applications for Euclid, Roman, and Rubin LSST}

\author[0000-0001-8019-6661]{Marziye jafariyazani}
\affiliation{Caltech/IPAC, 1200 E. California Blvd. Pasadena, CA 91125, USA}

\author{Daniel Masters}
\affiliation{Caltech/IPAC, 1200 E. California Blvd. Pasadena, CA 91125, USA}

\author{Andreas L. Faisst}
\affiliation{Caltech/IPAC, 1200 E. California Blvd. Pasadena, CA 91125, USA}

\author{Harry I. Teplitz}
\affiliation{Caltech/IPAC, 1200 E. California Blvd. Pasadena, CA 91125, USA}

\author{Olivier Ilbert}
\affiliation{Aix-Marseille Université, CNRS, CNES, LAM, Marseille, France}



\begin{abstract}
Entering the era of large-scale galaxy surveys which will deliver unprecedented amounts of photometric and spectroscopic data, there is a growing need for more efficient, data driven, and less model-dependent techniques to analyze spectral energy distribution of galaxies. In this work, we demonstrate that by taking advantage of manifold learning approaches, we can estimate spectroscopic features of large samples of galaxies from their broadband photometry when spectroscopy is available only for a fraction of the sample. This will be done by applying the Self Organizing Map (SOM) algorithm on broadband colors of galaxies and mapping partially available spectroscopic information into the trained maps. In this pilot study, we focus on estimating 4000\AA \space break in a magnitude-limited sample of galaxies in the COSMOS field. We use observed galaxy colors (ugrizYJH) as well as spectroscopic measurements for a fraction of the sample from LEGA-C and zCOSMOS spectroscopic surveys to estimate this feature for our parent photometric sample. We recover the D4000 feature for galaxies which only have broadband colors with uncertainties about twice of the uncertainty of the employed spectroscopic surveys. Using these measurements we observe a positive correlation between D4000 and stellar mass of the galaxies in our sample with weaker D4000 features for higher redshift galaxies at fixed stellar masses. These can be explained with downsizing scenario for the formation of galaxies and the decrease in their specific star formation rate as well as the aging of their stellar populations over this time period.


\end{abstract}



\section{Introduction} \label{sec:intro}

Measuring physical properties of galaxies is the first and key step to understand how galaxies form and evolve. One of the most common and least expensive ways to infer physical properties such as redshift, stellar mass and star formation rate (SFR) for statistically large numbers of galaxies over cosmic time is by measuring their energy output in multiple photometric filters. Multi-band photometric measurements provide data to construct the Spectral Energy Distribution (SED) of galaxies and then SED-fitting techniques can be employed to fit theoretical and empirical models to observed SEDs to infer multiple physical properties including stellar mass, SFR, age, reddening simultaneously, though with some limitations and known degeneracies (\cite{noll2009,walcher2011,Conroy2013,lower2020,Pacifici2023}).

On the other hand, high resolution spectra from galaxies enables the measurement of detailed characteristics of galaxies more robustly, including details of their chemical content and their star formation history, albeit at a much higher cost both at observing and data analysis stages. Therefore, the combination of wide multi-band photometric surveys observing millions of galaxies, with more restricted spectroscopic surveys targeting thousands of galaxies is the current scheme of galaxy observing programs aiming to measure physical properties of galaxies.

We are now entering the era of large scale galaxy surveys which are going to deliver unprecedented amounts of photometric and spectroscopic data. Euclid \citep{Euclid2011,euclid2013}, the Nancy Grace Roman Space Telescope \citep{wfirst2015,roman2019} and the Vera C. Rubin Observatory \citep{lsst2009,lsst2019} will each survey large areas of sky and provide multi-band photometric data for billions of galaxies which needs to be analyzed in more efficient ways. SED-fitting techniques as classical methods of inferring physical properties of galaxies from broadband photometry will be computationally expensive for these large datasets. Dependence of SED-fitting results on our current models and assumptions will be another limiting factor on properly analyzing new observations. On the other hand, spectroscopic observations will still be limited to only smaller subsamples of galaxies observed photometrically. Therefore, the need for more efficient, data driven, and less model-dependent techniques to analyze SEDs of galaxies to infer the most information out of them is becoming more crucial. A recent analysis by \cite{Nersesian2023} evidently showed that relying only on SED-fitting techniques to recover spectral features of galaxies from their broadband colors is not sufficient, and incorporating techniques to inform photometric analysis with high quality spectroscopy is necessary. Developing techniques to allow the integration of valuable but scarce spectroscopic information with broadband photometry will maximize the efficiency of both photometric and spectroscopic surveys, facilitating estimation of more reliable properties of galaxies. 

In this context, and with the expectation of high-dimensional data soon to become available for billions of galaxies, here we aim to make use of Manifold Learning as a data-driven approach to recover fundamental properties of galaxies from their broadband colors more efficiently. Manifold learning is a class of unsupervised algorithms that seek to learn the high-dimensional structure of the data without predefined classifications in order to find a lower-dimensional representation of that data which would be easier to visualize and interpret. This approach has been implemented in a handful of algorithms, each with their own capabilities and limitations, including Isometric Mapping (Isomap, \cite{isomap}), Self Organizing Maps (SOM, \cite{kohonen1982}), Locally Linear Embedding \citep{llembed}, t-distributed Stochastic Neighbor Embedding (t-SNE, \cite{tsne}) and Uniform Manifold Approximation and Projection (UMAP, \cite{umap}). The application of some of these algorithms in astronomical research has demonstrated great promise in analyzing photometric and spectroscopic data of stars, quasars and galaxies (\citealt{LLEexample2011,isomapexample2014,masters2015,LLEexample2017,tsneexample2017,tsneexample2018,Faisst2019,hemmati2019,tsneexample2019,tsneexample2020,tsneexample2020_2,LLEexample2021,umapexample2021,umapexample2021_2,Davidzon2022,chartab2023,umapexample2023}).

In this paper, we focus on using Self Organizing Map algorithm, which has been previously used for classification of light curves \citep{Brett2004}, estimating photometric redshift of galaxies (\cite{Way2012,kind2014,Greisel2015}) and finding and classifying variable Active Galactic Nuclei (AGNs, \cite{Faisst2019}). More recently, \citet{masters2019} used these maps as a tool to develop optimal photometric redshift calibration strategies for upcoming cosmology surveys with Euclid and Roman. In another work, \citet{hemmati2019} used this tool for both visualizing and computationally accelerating the estimation of galaxy physical properties like stellar mass and star formation rate (SFR) from photometric data by training SOMs with theoretical models, demonstrating that their method is able to estimate the full probability distribution functions for each galaxy significantly faster than traditional SED-modeling codes. Furthermore, \citet{Teimoorinia2022} use SOMs to classify spectra of galaxies in MaNGA survey \citep{Bundy2015}. They show that the distribution of spectra onto a SOM for a single galaxy can reveal the presence of distinct stellar populations within the galaxy, and therefore can be an alternative method to full spectral fitting for deriving the full probability distribution of physical properties, with the advantage of being more efficient than Bayesian methods. 

In this work, we advance one step further to use SOMs to estimate specific spectroscopic features of galaxies based on their broadband colors. This will be done by informing SOMs trained with a set of broadband colors with spectroscopic measurements for a sub-sample of the galaxies, and then estimate that feature for the rest of the training sample. In this paper, we specifically focus on estimating 4000 \AA \space break, which is a spectral feature of galaxies caused by absorption of metals in stellar atmospheres, indicative of lack of young stars and abundance of metal absorbents (\citealt{Hamilton1985,balogh1999}). It is therefore a powerful classical probe of age and metallicity of galaxies, and its relation with other properties of galaxies can put further constraint on galaxy formation models, including downsizing scenario \citep{Cowie1996}. 

This paper is organized as follows: In \S 2 we present the parent samples for our training data. \S 3 describes the algorithm we use and the training process. In \S 4 we use our method to predict D4000 feature for our photometric sample, and we study the D4000-Stellar Mass relation for our sample in \S 5. We finally discuss our results in \S 6. Throughout this work, we assume a $\Lambda$CDM cosmology with $H_0 = 70\,{\rm km\,s^{-1}\,Mpc^{-1}}$, $\Omega_\Lambda = 0.7$, and $\Omega_{\rm m} = 0.3$. All magnitudes are given in the AB system \citep{oke1983}. 
\pagebreak

\section{Sample} \label{sec:sample}
\subsection{Photometric Sample} \label{subsec:photometric}
We choose the training sample for this work from the latest data release of The Cosmic Evolution Survey (COSMOS, \citealt{scoville2007}). COSMOS started with HST/ACS observation of a 1.7 $deg^2$ mosaic reaching a depth of 27.2 AB in the F814W filter and then has been observed with most of major ground and space based telescopes to become a field with invaluable amount of both photometric and spectroscopic data. The new COSMOS2020 catalog (\citealt{cosmos2020}) includes multiband photometry in UV, optical and IR for about 1.7 million sources. 

\subsection{Spectroscopic Sample} \label{subsec:spectroscopy}
The spectroscopic measurements used in this paper come from two magnitude-limited surveys which measured specific spectroscopic features for galaxies in addition to their redshift in the COSMOS field: LEGA-C Survey (\citealt{legac2016}), and zCOSMOS Survey (\citealt{zcosmos2007}):

\begin{itemize}

  \item \textbf{LEGA-C}: The Large Early Galaxy Census (LEGA-C) is a spectroscopic survey of about 3200 K-band selected galaxies at redshift $0.6 < z < 1.0$ which is conducted with VIMOS/VLT (\citealt{legac2016}). This survey has a redshift-dependent magnitude limit of K$_{AB}$ = 20.7 - 7.5 $\times$ log((1+z)/1.8) which resulted in targeting only bright galaxies to obtain high quality spectra.
  
  \item \textbf{zCOSMOS}: zCOSMOS is a redshift survey in the COSMOS field conducted with VIMOS/VLT (\citealt{zcosmos2007}). For this work, we use zCOSMOS-bright data which is an I-band limited (I$_{AB}<22.5$) survey of about 20000 galaxies covering the whole COSMOS ACS field. The details of the spectral measurements for this sample of galaxies at $0 < z < 1.4$ can be found in \citealt{Vergani2010}.

\end{itemize}

\begin{figure*}
\includegraphics[width=\textwidth]{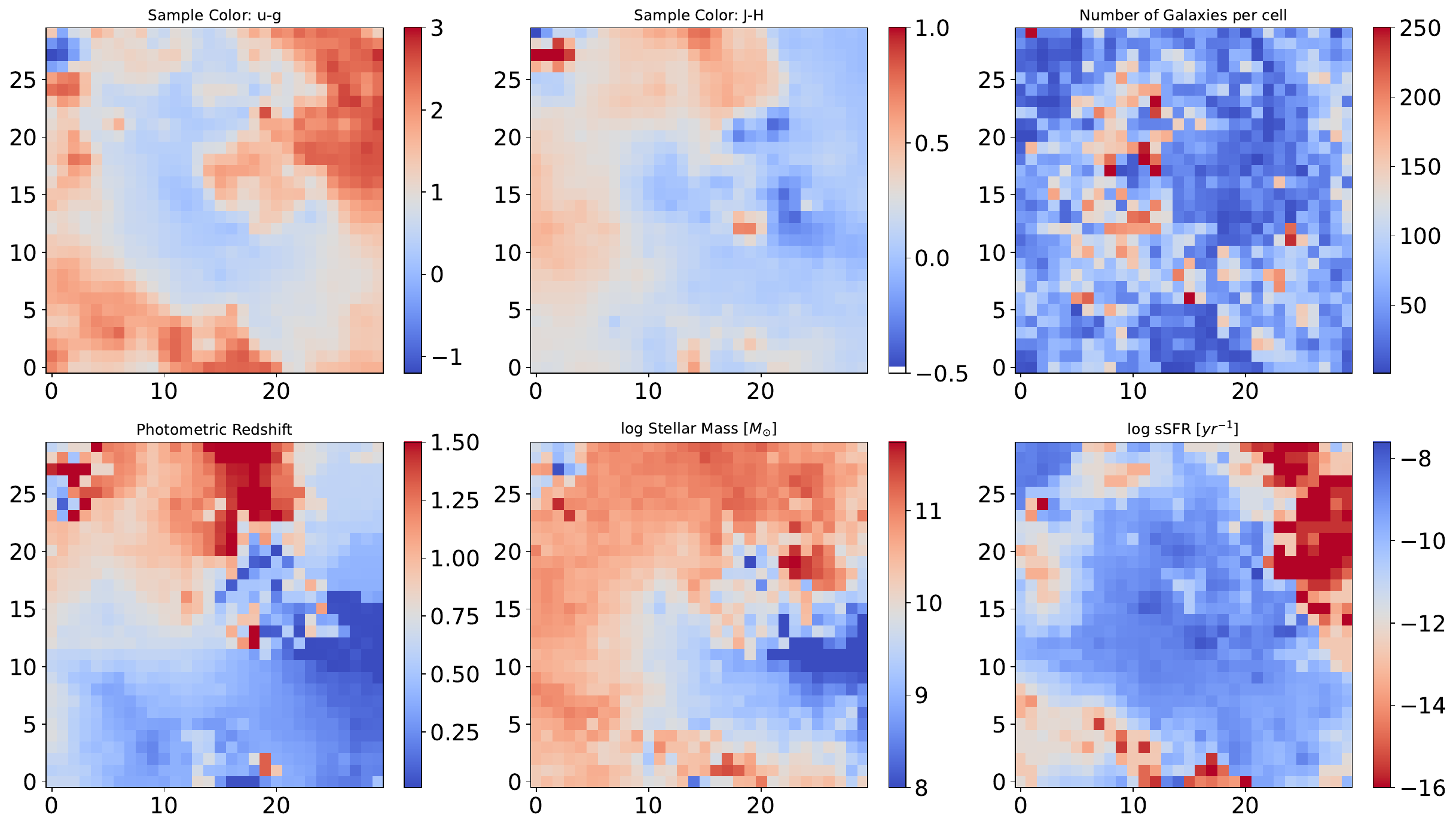}
\caption{Trained Self Organized Maps with 7 broadband colors (ugrizYJH) for our sample of 60511 galaxies. The top two left panels present two sample colors (u-g $\&$ J-H) on our trained maps. The top right panel is the occupation map showing the number of galaxies assigned to each cell during the training step. Bottom panels present physical properties of the trained sample measured using SED-fitting including photometric redshift, stellar mass and specific star formation rate. The value of each cell in these maps represent the median value of that property for the galaxies in the cell, adopted from COSMOS2020 catalogue.}
\label{fig:sample_color}
\end{figure*}

\begin{figure*}
\includegraphics[width=1\textwidth]{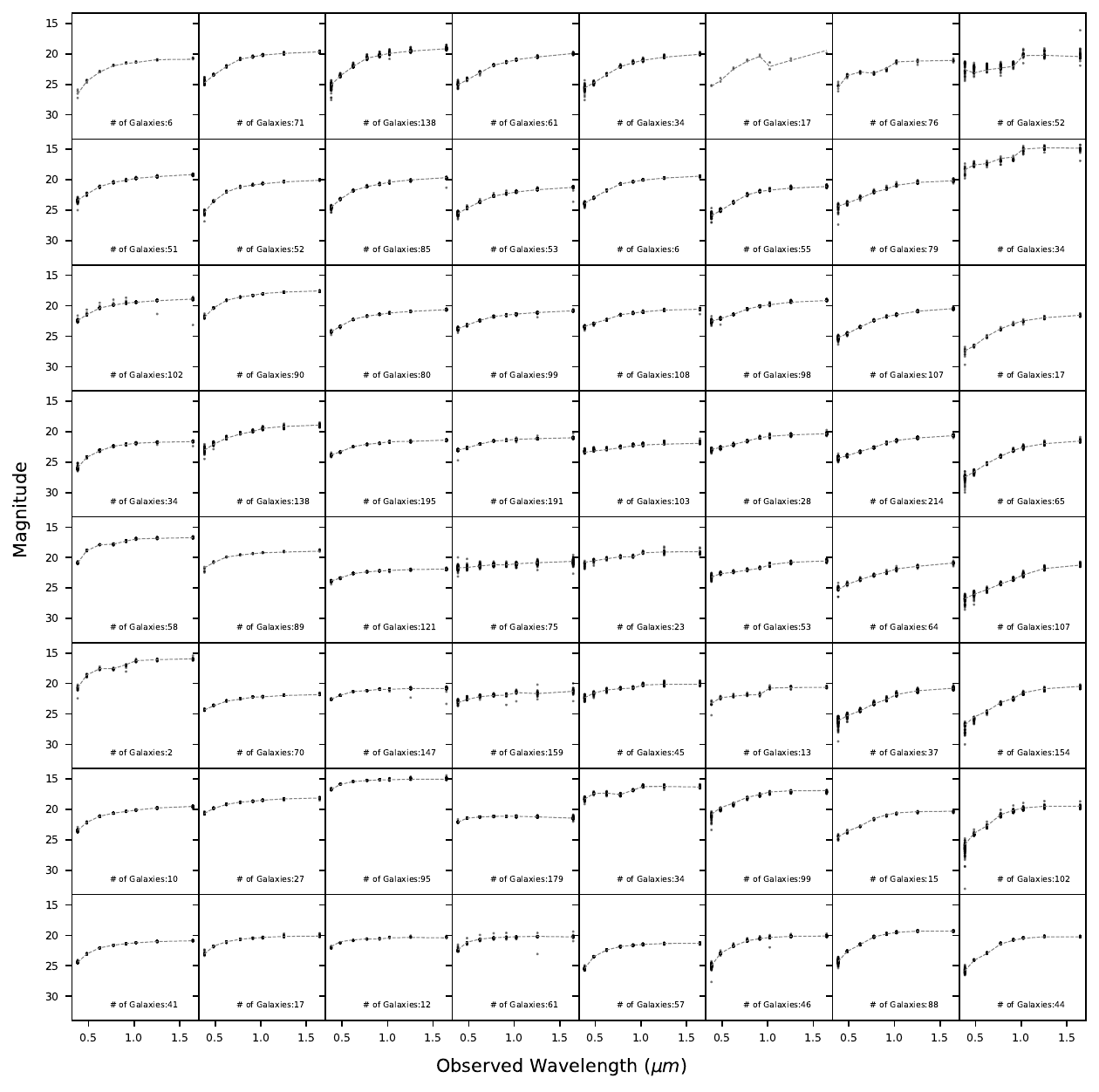}
\caption{This plot demonstrates SEDs of galaxies on trained SOMs. For visualization purposes, every one out of four cells in each row of the original map is shown in the plot. The median SED of each cell (dashed line), and the number of galaxies in each cell is shown on each panel.}
\label{fig:seds}
\end{figure*}

\section{Manifold Learning on broadband colors of Galaxies}
\subsection{Self Organizing Map Algorithm in a Nutshell} \label{sec:algorithm}

SOMs (\citealt{kohonen1982}) are a type of artificial neural network trained using unsupervised learning which are mainly used for dimensionality reduction as well as data visualization and clustering. The algorithm preserves the topological relationship of the input dataset which helps reveal correlations that are not easily identifiable in a high dimensional space. SOMs are usually used to bring down the dimension of a dataset to two for visualization purposes, though the technique can be used to reduce dimension to any arbitrary number.

This neural network algorithm consists of two layers of input and output data without any hidden layers in between. The input training data is a set of vectors of \textbf{N} dimensions, and the data is a feature map; a lattice of neurons (nodes) with topological positions (e.g. X,Y) where each neuron is assigned a weight vector with the same dimensionality as the input data. In the training step, the weight vectors are initialized randomly (or by sampling from the input data), and then SOMs would be trained by computing distance (e.g. Euclidean distance) between each neuron (neuron from the output layer) and the input data, and the neuron with the lowest distance will be the winner of the competition, known as Best Matching Unit (BMU). After finding the BMU, the value of that neuron and its neighbors will be updated, and the steps would be repeated until the maps reach convergence and datapoints with similar features end up in neighboring positions on the map. More details about this algorithm and its applications can be found in \cite{Yin2008} and \cite{cottrell}.

\begin{figure}
\includegraphics[width=0.45\textwidth]{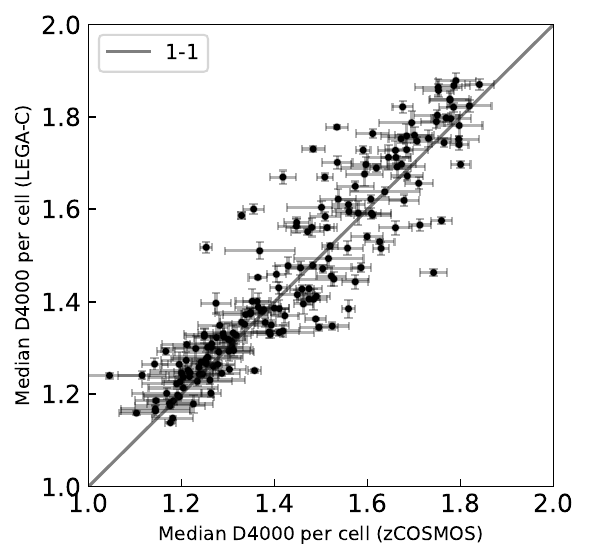}
\caption{This plot compares median D4000 measured from spectra taken in LEGA-C and zCOSMOS surveys for galaxies in each cell of our trained SOM. Only cells which have at least three D4000 measurements in each survey are shown here. The plot shows a general consistency between the measurements in LEGA-C and zCOSMOS.}
\label{fig:cell_comparison}
\end{figure}

\subsection{Training SOMs} \label{sec:training}
We train our SOMs with 7 broadband colors derived from u, g, r, i, z, Y, J and H filters for a subsmaple of 60511 galaxies in the COSMOS field. We select these galaxies from COSMOS2020 photometric catalog such that they meet the magnitude criteria for either zCOSMOS (I$_{AB} <$ 22.5) or LEGA-C survey (K$_{AB} <$ 21.08 at $z \sim$ 0.6), with an additional condition that they should have at least four measured colors in our selected filter set. This specific filter set is analogous to what will soon become available for a statistically large sample of galaxies thanks to Euclid and Vera Rubin telescopes. The NISP instrument onboard Euclid will provide us with broadband photometry in Y, J and H filters for about 15,000 $deg^2$ of the sky where approximately 9000 $deg^2$ of the region would also be observed with Vera Rubin telescope providing photometry in u, g, r, i and z filters. 

These trained SOMs are presented in Fig. \ref{fig:sample_color} where top left panels show two sample colors (u-g and J-H) organized on the maps. As expected, galaxies are clustered on these maps according to their colors. The top right panel of the same figure shows an occupation map where each cell of the map is color-coded based on the number of galaxies contained in that cell which ranges from 1 to 380 galaxies with a median of 56 galaxies per cell. This classification based on colors suggests that these galaxies are also organized in terms of their physical characteristics on the map. To examine this assumption, we color-code these maps with some physical features that were not fed into the algorithm during the training step. Here, we use the physical properties derived from SED-fitting for these galaxies available in COSMOS 2020 catalog (\citealt{cosmos2020}). These are shown in the bottom panels of the Fig.\ref{fig:sample_color}. The lower left panel presents photometric redshift, where the value of each cell is the median photometric redshift of the galaxies in that cell. The middle and right panels have been calculated in the same way for stellar mass and specific star formation rate (sSFR) of these galaxies. These three maps visually demonstrate how training SOMs with broadband colors leads to classifying galaxies based on their fundamental physical properties without any modeling assumptions. 

It should also be noted that the physical properties mapped on to these SOMs are the results of SED-fitting on up to 31 photometric bands available in the COSMOS2020 catalog, while we only used 8 bands in training our SOMs. These maps suggest that the groups of galaxies in the cells where dispersion of their colors are small (i.e. have similar SEDs), potentially have very similar physical properties allowing us to estimate further features that have not been explicitly measured for all the galaxies in that cell. Furthermore, measuring absolute physical properties is attainable as the color-based trained SOMs has already separated galaxies in the redshift space.

Our 30 $\times$ 30 maps categorized our $\sim$ 60000 galaxies into 900 classes. This size for our training map is determined in an iterative process such that individual cells are neither underpopulated nor overpopulated to cause high dispersion in galaxies' colors. Fig. \ref{fig:seds} presents individual SEDs of galaxies in 64 cells of our map corresponding to every four cells of the original 30 $\times$ 30 map. The number of galaxies, and the median SED of each cell (dashed line) is also presented in each panel. It should be emphasized while galaxies within the same cell have the highest degree of similarity, neighboring cells also have more similar properties compared to galaxies in distant regions of the map. By taking advantage of this model-free SED grouping, we then use physical features measured for a sub-sample of galaxies in each cell to predict that feature for the rest of the galaxies in that cell. To investigate this approach and its potentials, we estimate 4000\AA \space break for our sample in Sec \ref{sec:4000} which classically requires spectroscopy or narrow-band photometry for proper estimation (\cite{Stothert2018,Renard2022}).

\section{Predicting 4000\AA \space break from broadband photometry} \label{sec:4000}
4000\AA \space break, a probe of age and metallicity of galaxies, is quantified by measuring the ratio between the continuum fluxes on the red and blue side of this break. D4000 is one of the features that have been measured in both LEGA-C and zCOSMOS surveys for a statistically large number of galaxies in the COSMOS field, as it can be measured in lower quality spectra compared to the higher resolutions needed to measure individual spectral lines. In this work, we use D4000 measurements for 2241 galaxies from LEGA-C survey and 10622 galaxies from zCOSMOS survey, giving us the proper statistics for this pilot study.

In this paper, we adopt the definition of \citealt{balogh1999} ($D4000_{\mathrm{n}}$) for the strength of 4000\AA \space break as it is less sensitive to reddening effects compared to the original definition of the D4000 ($D4000_{\mathrm{w}}$, \citealt{bruzual1983}) which uses a larger wavelength range to define the continuum around the break. We refer to D4000$_{\mathrm{n}}$ by simply D4000 throughout this paper.

\begin{equation}
D4000_{\mathrm{n}}=\frac{F(4000-4100 \ \AA)}{F(3850-3950\ \AA)}
\label{eq:d4000} 
\end{equation} 

\begin{figure*}
\includegraphics[width=\textwidth]{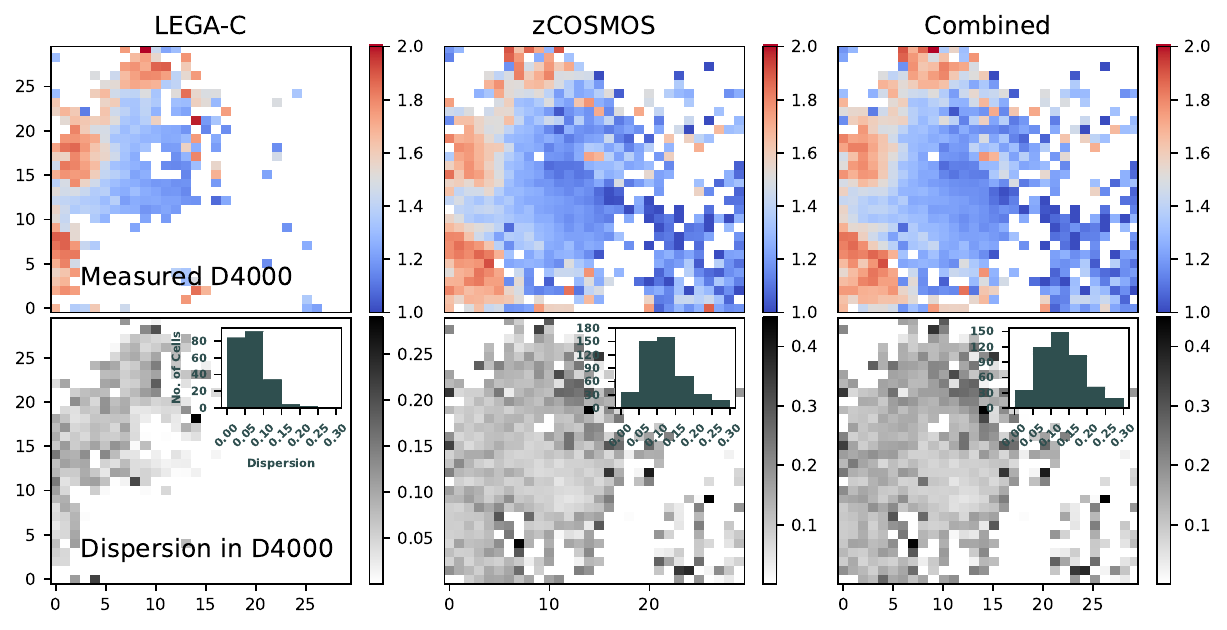}
\caption{The top row of this plot presents D4000 measurements available for a fraction of our sample mapped on the trained self organized maps measured from LEGA-C survey (left), zCOSMOS survey (middle), and the median of the measurements from both surveys (right). The value for each cell is the median of the measured D4000 for the galaxies residing in that cell where at least 3 spectroscopic measurements exist. Bottom row shows the dispersion in D4000 for every cell of the SOM. The inset histogram in each panel presents the distribution of the median dispersion among cells. zCOSMOS survey includes lower mass and fainter galaxies with more uncertain measurements and therefore higher dispersion is seen for estimations from this survey.}
\label{fig:properties}
\end{figure*}

In the first step, we map D4000 measurements from these two surveys on our trained SOMs described in Sec. \ref{sec:training}. 
We then compare these measurements from the two surveys for the galaxies in our sample. For this purpose, we use the median D4000 measurements for the galaxies lying in each cell of the trained SOM from LEGA-C and zCOSMOS as shown in Fig \ref{fig:cell_comparison}. There are 282 number of cells in our trained maps which have at least one galaxy with measured D4000 in the LEGA-C survey ($31\%$ of the cells), and 617 cells with at least one measurement in zCOSMOS survey ($69\%$ of the cells). Fig \ref{fig:cell_comparison} compares the median of these measurements for cells which have at least 3 measurements in each survey showing the consistency between measurements in two surveys with a tighter relation at lower D4000 values.

Distribution of these spectroscopic measurements on our trained SOM is presented in Fig. \ref{fig:properties}. In the top row, the left panel represents median D4000 measurements for the sub-sample of our galaxies observed in LEGA-C, middle panel is from zCOSMOS, and the right panel includes D4000 measurements from both surveys. These panels reveal two distinct populations of galaxies with weak and strong D4000 features, blue and red respectively. Comparing these maps with sSFR map (lower right panel of Fig. \ref{fig:sample_color}) demonstrates that, as we expected, galaxies with lower sSFR have stronger D4000 feature, and galaxies with high sSFR and abundance of younger stars have weaker D4000. The data from the zCOSMOS Survey (middle panel of Fig. \ref{fig:properties}) has more coverage on the D4000 map as it includes fainter and lower mass galaxies in a wider redshift range compared to LEGA-C. The right panel of Fig. \ref{fig:properties} includes all the D4000 measurements available from these two spectroscopic surveys for our training sample, which will be used to estimate this feature for the rest of our galaxies. 

The bottom row of Fig. \ref{fig:properties} presents the dispersion map of D4000 measurements in each cell of the SOM. The inset plots in the bottom row present the distribution of the cells in terms of the amount of the D4000 dispersion. For the LEGA-C map, most of the cells ($81\%$) have a dispersion of less than 0.1 and for the zCOSMOS map, most of the cells ($72\%$) have dispersion of less than 0.15. Also, cells with at least three D4000 measurements have a median sigma value of 0.12 which does not decrease significantly with more number of spectroscopically-measured D4000 in the cell. Therefore, we evaluate the potential use of the median value of the spectroscopic D4000 measurements to estimate this feature for the rest of the galaxies in that cell with no available spectroscopy, where at least three spectroscopic measurements exist in their host cell.

\begin{figure*}
\includegraphics[width=\textwidth]{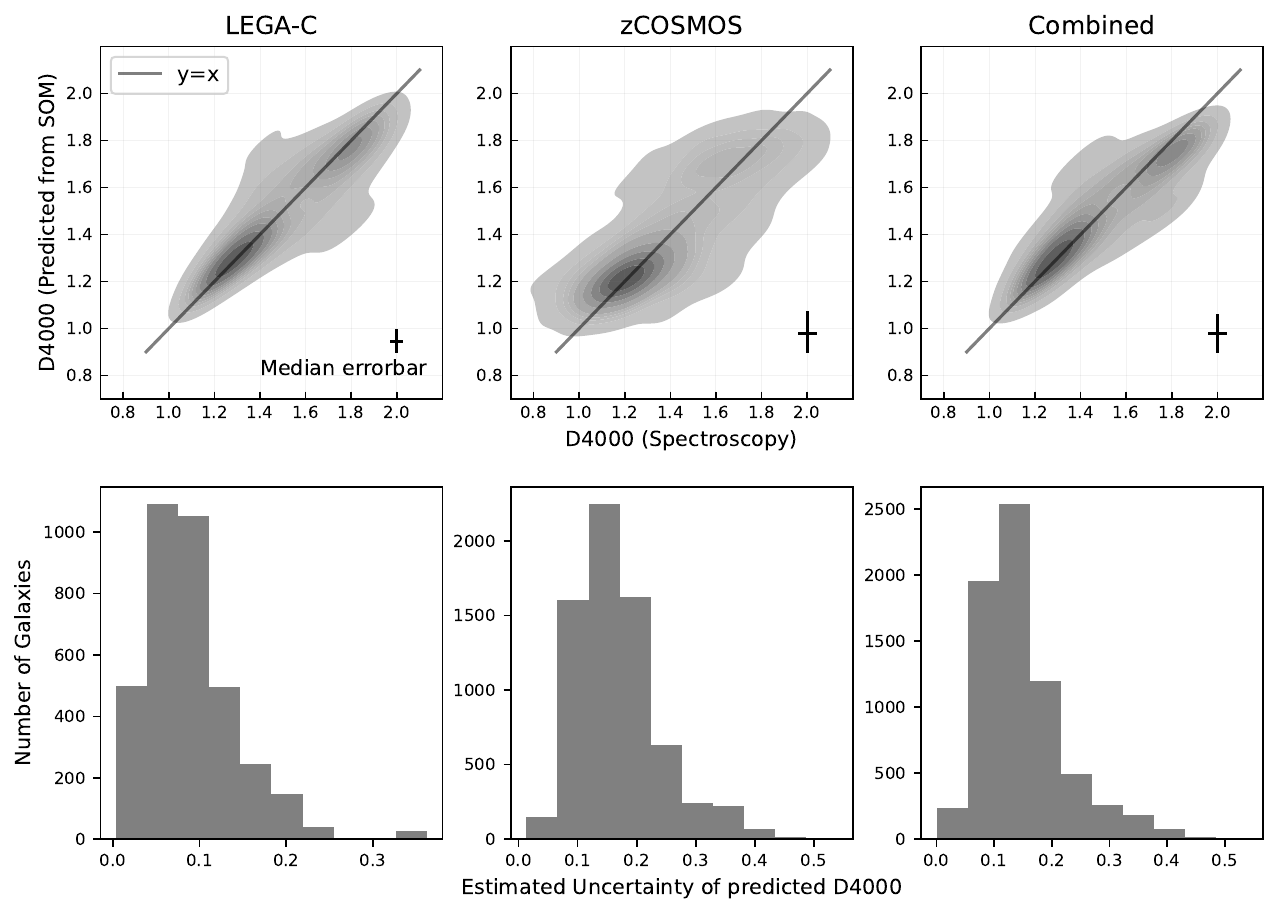}
\caption{Top panels present predicted D4000 for 20$\%$ of the sample based on data of rest of 80$\%$ of the sample as a function of the true value of D4000 measured in spectroscopy. Left panel is predictions based on LEGA-C only, the middle panel is for zCOSMOS only, and the right panel is based on both surveys combined. Median error bar of all the measurements is shown on the lower-right corner of panels. D4000 for blue galaxies which dominate the sample are best estimates as we can see on these plots. Bottom panels show the histogram of galaxies based on the uncertainty of their D4000 predictions (combination of spectroscopic  measurement errors and dispersion of D4000 in each cell of SOM) based on each survey. Estimates based on the LEGA-C survey are more accurate as it has lower measurement error and includes a less diverse population of galaxies.}
\label{fig:prediction}
\end{figure*}

To evaluate this approach, we use 80$\%$ of our spectroscopic data to estimate the D4000 features for cells with at least three measurements. We then compare our estimated values and their uncertainty with the actual D4000 measurement from spectroscopy for the remaining 20$\%$ of the spectroscopic sample. The top panels of Figure \ref{fig:prediction} show this comparison where predicted D4000 for 20$\%$ of the sample based on the trained SOM with data for the 80$\%$ of the sample is presented as a function of D4000 measured in spectroscopic surveys for the same galaxies. From left to right, each panel shows estimation based on LEGA-C data, zCOSMOS data and the combination of both surveys. The bottom panels of Fig.\ref{fig:prediction} present the distribution of uncertainties of the predicted D4000 values for each case. As expected from different depth and measurement errors of LEGA-C and zCOSMOS surveys, which significantly affect the total uncertainty, estimations based on LEGA-C which targeted brighter galaxies with lower measurement errors have lower uncertainties with the peak of uncertainty distribution at $\sim$ 0.08, whereas estimations based on zCOSMOS which includes fainter galaxies with higher measurement errors are more uncertain with the peak of distribution at $\sim$ 0.16. Comparing these uncertainties with the median spectroscopic measurement errors for this sample in the two surveys ($\sim$ 0.05 for LEGA-C, and $\sim$ 0.08 for zCOSMOS) suggests that our method, with the number of spectroscopic measurements we have, can recover D4000 feature based on broadband SEDs of galaxies without any modeling assumption, with an uncertainty in the order of twice the uncertainty of the measurement error of the employed spectroscopic data. It should be noted that for areas of the map that we do not have enough D4000 measurements from spectroscopy, specifically for cells without any measurements in their nearby neighbors as well, we do not constrain the D4000 feature, and galaxies in these cells can be targets of great interest for future spectroscopic follow-ups. 

\begin{figure}
\includegraphics[width=0.47\textwidth]{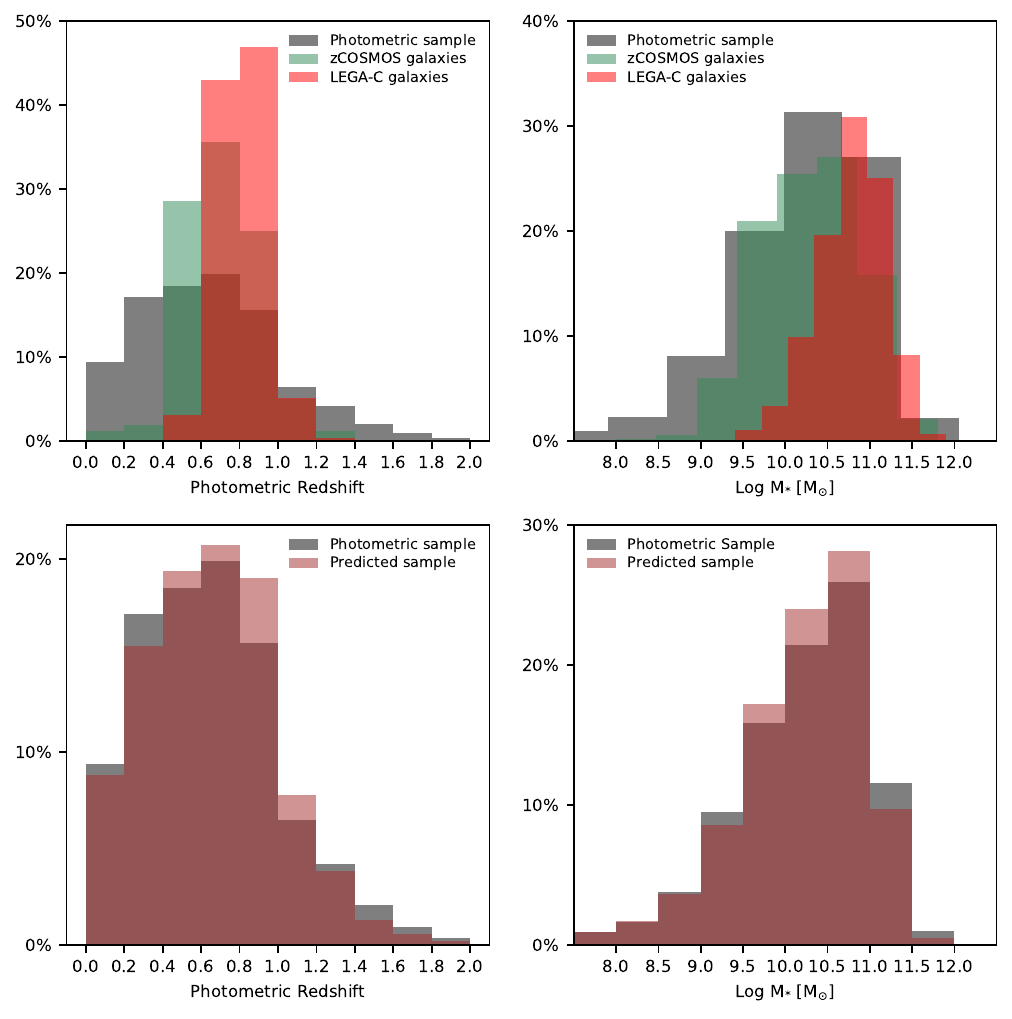}
\caption{The top left panel presents the redshift distributions of our magnitude-limited parent photometric sample compared to our spectroscopic samples from LEGA-C and zCOSMOS surveys. The top right panel is the stellar mass distributions of these three samples. The bottom panels present the same distributions for the photometric sample compared to predicted sample, which is $\sim 79\%$ of galaxies in our photometric sample that we are able to predict their D4000.}
\label{fig:histogram}
\end{figure}

To compare the distribution of the galaxies with spectroscopic measurements with galaxies that we are able to predict their D4000 feature and our parent photometric sample, we then look at the distributions of the redshift and stellar mass of these samples. The top left panels of Fig.\ref{fig:histogram} present the redshift distributions of our parent photometric sample, LEGA-C galaxies with measured D4000 and zCOSMOS galaxies with measured D4000, and the top right panel is the stellar mass distributions of these three samples. This is to demonstrate that our magnitude-limited photometric sample has an extended range of redshift and stellar mass compared to the employed spectroscopic samples. The bottom two panels of Fig.\ref{fig:histogram} present the distributions for the photometric sample and predicted sample, that is the galaxies that we are able to predict their D4000, which is $\sim 79\%$ of our photometric sample. The bottom panels show that our predicted sample has about the similar redshift and stellar mass distribution compared to the parent sample extending our estimations beyond the redshift and stellar mass limits of our employed spectroscopic samples.

\begin{figure}
\includegraphics[width=0.45\textwidth]{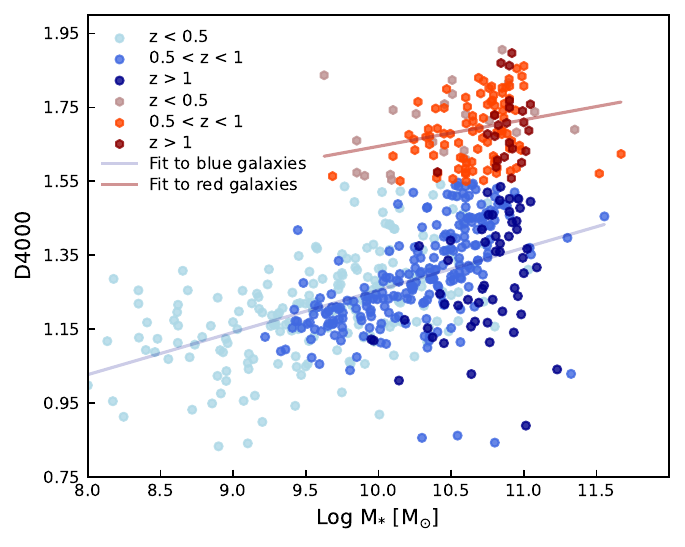}
\caption{D4000 as a function of stellar mass based on estimated 4000\AA \space break from SOMs where each datapoint represents median D4000 and stellar mass of a cell. Galaxies are divided in two populations of blue (D4000$<$1.55) and red sources (D4000$>$1.55) in three redshift bins of z $<$ 0.5, 0.5 $<$ z $<$ 1 and z $>$ 1. A linear fit to the blue and red populations at all redshifts is also presented on the plot.}
\label{fig:mass-d4000}
\end{figure}

\section{D4000-Mass Relation} \label{sec:d4000-mass}

In this section, we use our predicted D4000 features based on SOMs to study the D4000-Mass relation for the galaxies in our sample. This relation has been studied in the literature based on D4000 features from spectroscopy and narrow-band photometry (\cite{Haines2017,Siudek2017,kim2018,Renard2022}). These studies observed a bimodal distribution of galaxies on the D4000-Stellar Mass plane where they found blue cloud and red sequence galaxies differentiated from each other. 
Here, we use the estimated stellar mass of each galaxy based on SED-fitting available in the COSMOS2020 catalog. Fig.\ref{fig:mass-d4000} presents the D4000-stellar mass relation for the galaxies in our sample, where each datapoint represents an individual cell of SOM with median properties of galaxies lying on that cell. We divide our galaxies in three redshift bins of z $<$ 0.5, 0.5 $<$ z $<$ 1 and z $>$ 1. Overall, we observe a positive correlation between the strength of 4000\AA \space break and stellar mass of galaxies. This behavior can be explained with downsizing scenario for the formation history of galaxies where stars formed earlier in more massive galaxies during intense star formation episodes and therefore producing stronger 4000\AA \space break(\cite{Cowie1996,DeLucia2006,gonzalez2008,Pacifici2016}). Also, galaxies at the lowest redshift bin tend to have stronger break compared to the higher redshift galaxies at fixed stellar masses. The increase in D4000 over time could be explained by the overall aging of stellar populations of galaxies as well as the decrease in specific star formation rate of main sequence of galaxies over the past $\sim$10 billion years (\cite{Noeske2007,Ilbert2015}.

We then divide our sample into two populations based on their D4000 as late-type (blue, D4000$<$1.55), and early-type (red, D4000$>$1.55) galaxies following \cite{Kauffmann2003}. Fig.\ref{fig:mass-d4000} demonstrates that dependence of the strength of D4000 on stellar mass of galaxies is slightly shallower for red galaxies with a slope of 0.07, compared to the slope of 0.12 for blue galaxies. The D4000-stellar mass relation measured for our blue galaxies is comparable to the relations reported in \cite{Haines2017} who measured a slope of 0.12-0.14 for D4000-mass relation for their blue galaxies with little to no redshift dependence at 0.5 $< \textit{z} <$1.1, based on data for $\sim$ 65000 galaxies from the VIPERS survey (VIMOS Public Extragalactic Redshift Survey, \cite{Guzzo2014}) . \cite{Renard2022} also measured a slope of $\sim$ 0.09-0.12 for blue clouds based on D4000 measured from narrow-band photometry for $\sim$17000 galaxies at 0.5 $< \textit{z} <$ 1 in PAUS Survey (Physics of the Accelerating Universe Survey, \cite{ben2009}). However, both of these studies measured a steeper slope for their red-sequence galaxies, $\sim$0.23 in \cite{Haines2017} and 0.25-0.32 in \cite{Renard2022}. One potential source of larger discrepancy in the measured relation for the red population can rise from the wider redshift range of our sample as well as the fact that these samples are magnitude limited and therefore are not mass complete, leaving us with most massive and bluer galaxies at higher redshift bins and low fraction of red galaxies in general. For instance, only $\sim$20$\%$ of our galaxies and 35$\%$ of the sample in \cite{Renard2022} meet the criteria for red sequence in each study.

\section{conclusions} \label{sec:discussion}
This work demonstrates the power of manifold learning to predict spectroscopic features of statistically large samples of galaxies when we have their broadband colors as well as spectroscopy only for a fraction of the sample. Here, we focused on the D4000 feature which was measured for statistically large number of galaxies over a wide redshift range in spectroscopic surveys, giving us the proper statistic we needed for this analysis. However, this method can be extended to predict any spectroscopic features like emission and absorption lines from broadband photometric surveys when we have a certain number of spectroscopic measurements to fill in any trained Self Organized Map. One of the key benefits of this method is that it is purely data-driven and is completely independent of any modeling assumption, unlike traditional methods like SED-fitting.

Another benefit of this approach is when we acquire minimum number of spectra to fill in the cells of our trained SOM, we will then be able to predict specific spectroscopic features for unlimited number of galaxies with observed colors with the only requirement that they meet the criteria of the magnitude limit of the spectroscopic data. The minimum number of required spectra will depend on the diversity of galaxies under investigation as well as the required precision in the estimation of any specific feature. For example, we showed that in this pilot study we are able to estimate D4000 feature for galaxies in cells hosting average of $\sim$ 67 galaxies when we have at least 3 spectroscopic measurements in that cell, and the uncertainty in our estimations do not change significantly with having more spectroscopy in each cell. Also, using this method, we can potentially identify cells filled with galaxies with little or no spectroscopic measurements which will be high-priority candidates for follow-up spectroscopy to build unbiased and complete galaxy samples for any science investigation.

Another promising potential of this method is to identify outlier galaxies. After the training step, distinct galaxies in individual SOM cells, cells with very distinct properties compared to their neighboring cells as well as cells which host very low numbers of galaxies can be flagged to be further analyzed as potential outlier sources. These outliers could be the result of data-processing issues, observational constraints such as blended sources, or can lead to discovery of intrinsically unique and rare sources. In each case, correctly identifying these sources has huge benefits both for improving data analysis pipelines and for addressing specific science questions that can transform our view of a specific subject. The routine to identify outlier sources with this method is now being developed within our team and will be published in a subsequent paper.

\vspace{5mm}




\typeout{}
\bibliography{ref}

\bibliographystyle{aasjournal}



\end{document}